\documentclass[referee,usenatbib]{mnras}
\usepackage{mathptmx}
\usepackage[T1]{fontenc}
\usepackage{ae,aecompl}
\usepackage{graphicx}	
\usepackage{amsmath}	
\usepackage{amssymb}	
\title[Cosmic Horseshoe gravitational lens]{Reconstructing the cosmic Horseshoe gravitational lens using the singular perturbative approach.}
\author[Alard, C.]{Alard, C., 
\\
IAP, 98bis Boulevard Arago, Paris \\}
\date{July 5th 2016}
\pubyear{2016}
\begin{document}
\label{firstpage}
\pagerange{\pageref{firstpage}--\pageref{lastpage}}
\maketitle
%
\begin{abstract}
The cosmic horseshoe gravitational lens is analyzed using the perturbative approach. 
The two first order perturbative fields are expanded in Fourier series. The source is reconstructed
using a fine adaptive grid. The expansion of the fields at order 2 produces a higher value of the chi-square.
 Expanding at order 3 provides a very significant improvement, while
order 4 does not bring a significant improvement over order 3. The presence of the order 3 terms is not a consequence of limiting
the perturbative expansion to the first order. The amplitude and signs of the third order terms are recovered by including the contribution
of the other group members. This analysis demonstrates that the fine details of the potential of the lens could be recovered
independently of any assumptions by using the perturbative approach.
\end{abstract}
\begin{keywords}
gravitational lensing: strong - methods: numerical
\end{keywords}
\section{Introduction}
Strong gravitational lensing offers a unique opportunity to probe the dark halos potential in the vicinity of the Einstein circle.
The lensing potential relates directly to the projected matter distribution and as a consequence is a direct measurement of the matter
distribution in the lens. However deriving a precise relation between the observations of a gravitational lens and the lensing potential is generally difficult.
There are basically two main problems. First the lens may show some degree of complexity and may not be properly described with
simple analytical models. And secondly some degree of degeneracy in the modeling of the lens is generally present see for instance, ~\cite{Saha2006},
~\cite{Wucknitz2002}, ~\cite{Chiba2002}. A solution to the first point is to use a non parametric method for the reconstruction of the potential.
For instance potential reconstruction on a grid offers a general model free solution, but the obvious drawback is a dramatic increase in the number
of parameters, which in turn aggravates the degeneracy issue. The only solution is to developp a method that offers a direct relation between
the arc morphology and the potential. This is precisely what the perturbative approach (~\cite{Alard2007}) achieves. At first order the potential
is expanded using two angular functionals, the fields $f_1$ and $\frac{d f_0}{d \theta}$. Each of these fields relates directly to the arc morphology,
$f_1$ is related to the mean radial position of the arc, while $\frac{d f_0}{d \theta}$ is related to width of the arc in the radial dimension.
This direct relation offers a simple solution to the degeneracy problem. Another aspect is that the reconstruction of the potential is general and does
not require any specific assumptions. Some specific examples of reconstruction of arcs systems using the perturbative approach are presented in
~\cite{Alard2009} and ~\cite{Alard2010}. In particular the reconstruction the lens in ~\cite{Alard2009} shows that very complex systems can be handled
in this approach. The application of the perturbative method to the cosmic horseshoe gravitational lens offers the possibility to push the reconstruction
to a high level of accuracy. The HST data available for this lens offer an excellent resolution and a wealth of details allowing to probe
the fine details of the halo dark matter distribution.  Let's recall
the basic equations of the order one theory by starting from the lens equation, Eq. (~\ref{lens_eq}).
\begin{equation}
{\bf r_S} = {\bf r} -\nabla \phi
\label{lens_eq}
\end{equation}
Using the equations relating the fields $f_0$ and $f_1$ to the potential,
\begin{equation}
\left\{
\begin{aligned}
\phi(r,\theta) &= \phi_0(r)+\epsilon \psi(r,\theta) \\
\psi(r,\theta) &= f_0(\theta)+f_1(\theta) (r-1)
\end{aligned}
\right.
\label{pot_def}
\end{equation}
The lens equation Eq. (~\ref{lens_eq}) is expanded to order one in $\epsilon$ (~\cite{Alard2007}):
\begin{equation}
 {\bf r_S} = \left(\kappa_2 \ dr-f_1 \right) {\bf u_r} - \frac{d f_0}{d \theta }\bf {u_{\theta}}
 \label{pert_0}
\end{equation}
With:
\begin{equation}
 f_1=\left[\frac{d \psi}{d r} \right]_{r=1}  \ \ ; \ \ f_0=\psi(1,\theta) \ \ ; \ \  \kappa_2=1-\frac{d^2 \phi_0}{d r^2}
\label{eq_f1_df0}
\end{equation}
It is useful to introduce the impact parameter of the source, namely ${\bf r_S=\tilde r_S+r_0}$ leading to:
\begin{equation}
 {\bf \tilde r_S} = \left(\kappa_2 \ dr-\tilde f_1 \right) {\bf u_r} - \frac{d \tilde f_0}{d \theta }\bf {u_{\theta}}
 \label{pert_1}
\end{equation}
With:
$$
 \tilde f_i=f_i+x_0 \cos(\theta)+ y_0 \sin(\theta) \ \ \ \ , i=0,1
$$
And the impact paremeter vector ${\bf r_0}$
$$
{\bf r_0}=(x_0,y_0)
$$
\section{Building the perturbative solution.}
In the perturbative approach (Alard 2007, Alard 2010, 2011) the construction of the solution
requires the evaluation of the two fields $f_1(\theta)$ and $\frac{d f_0}{d \theta}$. These
two quantities are functions of the angular variable in the lens plane $\theta$. We recall
that these two fields have simple interpretations for a strong gravitational lens.
For a circular source with radius $r_0$ the positions of the images contours are given by ~\cite{Alard2007}:
\begin{equation}
 \kappa_2 dr = f_1 \pm \sqrt{r_0^2-\frac{d f_0}{d \theta}^2}
\label{eq_circ}
\end{equation}
For convenience the Einstein radius is fixed to unity by using a proper choice of radial units, as a consequence,
$r=1+\epsilon dr$ with $\epsilon \ll 1$. It is straightforward to deduce from Eq. ~\ref{eq_circ} that $f_1$ corresponds to the mean position
of the two image contours and that image are formed in the angluar domain defined by: $\frac{d f_0}{d \theta} < r_0$.
It is also clear from Eq. ~\ref{eq_circ} that the maximum image width is obtained when $\frac{d f_0}{d \theta}=0$.
\subsection{Initial approximation}
The circular source model is used as a first guess in order to estimate the parameters of a piecewise polynomial model of the fields
. The reconstruction of $f_1$ is direct, while 
 $\frac{d f_0}{d \theta}$ is estimated using the following constraints: the field is near zero in the
central region of each images, and the the field is above a threshold value in the dark areas.
In this particular lens the nature of the first guess is simplified since the structure of the images
 reveal a fold configuration. The initial estimation of $f_1$ is done by taking the mean position of the
bright spots in the image. An estimation of $f_1$ can also be done by estimating the closest
fold configuration. These two estimates gives quite similar results and are both appropriate as a first
guess. A similar approach is used for $\frac{d f_0}{d \theta}$ one can reconstruct the field by estimating
its local behavior near the nodes and make an interpolation with the constraint that the field must
be above a certain threshold in the dark areas. As for
$f_1$ taking the closest fold configuration give quite similar results. This initial guess is consistent
with the topological properties of the solution and is an approximate description of its general shape.
To reach the optimal level of accuracy this solution needs to be refined numerically. The numerical refinement
could be performed directly on the polynomial piecewise elements, but since the Fourier expansion of the
fields is related to the multipole expansion of the potential, it is more efficient to expand the fields
using Fourier series. 
\begin{equation}
\left\{
\begin{aligned}
f_1 &=& \sum_n \alpha_{1,n} \cos(n \theta)+\beta_{1,n} \sin(n \theta) \\ 
\frac{d f_0}{d \theta} &=& \sum_n \alpha_{0,n} \cos(n \theta)+\beta_{0,n} \sin(n \theta)
\end{aligned}
\right.
\end{equation}
\subsection{Source and Image reconstruction.} 
The first step in the construction of the refined numerical model is to make an initial estimation
of the Fourier series coefficients. This is easily accomplished by computing the scalar product
of the Fourier basis function with the piecewise polynomial model. Once the Fourier coefficient
are initiated the optimal value of the coefficient is computed by minimizing the chi-square
between the re-constructed image of the source and the HST data. The reconstruction of the source
and its associated images are complex processes which we now describe in details.
The reconstruction of the source and of its associated images requires the use of finer grids.
A finer grid in the image plane is formed by sub-dividing the pixels. The typical resolution is
of a factor of 10 in each direction. The values of the image on the sub-grid is obtained
by B-spline interpolation. This sub-grid in the lens plane is transported in the source
plane using the perturbative lens equation at order one (Eq. ~\ref{pert_0}).
The size of the grid in the source plane is adapted to the resolution of the grid in the lens plane.
%
%
The grid size in the source plane has to be small enough to allow the reconstruction of all the image
details. Basically two image structures should not merge in the same pixel. Experiment are made to obtain
a suitable minimal value of the grid size. In some areas the grid size has to be adapted by extending the size
in order to have at least a few points from the image plane falling into the bin in the source plane. 
When several values fall in the
same cell of the source plane (which is by definition the case when several images are formed) the
different values are averaged. Once a source model is constructed the corresponding image is formed
by estimating the values of the pixels on the finer grid in the lens plane. The finer grid image is integrated
to produce an image at the initial resolution. 
\label{Source_rec}
\subsection{Image deconvolution.}
The source and image reconstruction procedure we described ignores the problem of the convolution of the image
with the PSF. A numerical model of the PSF is reconstructed using the Tiny Tim software (~\cite{Krist2011}). 
To correct for the effect of the PSF convolution on the image an iterative procedure 
is implemented. The difference image between the actual HST image and the image of the source is sent back
to the source plane. This correction is added to the initial source reconstruction and a new image is
computed. Once again the difference with the actual image is taken and the procedure is iterated. The convergence
of this method is fast and after a few iterations a convergence is achieved. Too many steps of this iterative
add only noise and as a consequence it is optimal to stop the iteration after a proper number of steps. Experiments
shows that the optimal number of steps is between 5 and 10. Finally a number of 8 steps was adopted, however
changing this number and taking any number between 5 an 10 does not produce significant changes.
\label{Deconv}
\subsection{Accurate numerical estimation of the fields.}
\begin{figure}
 \includegraphics[width=\columnwidth]{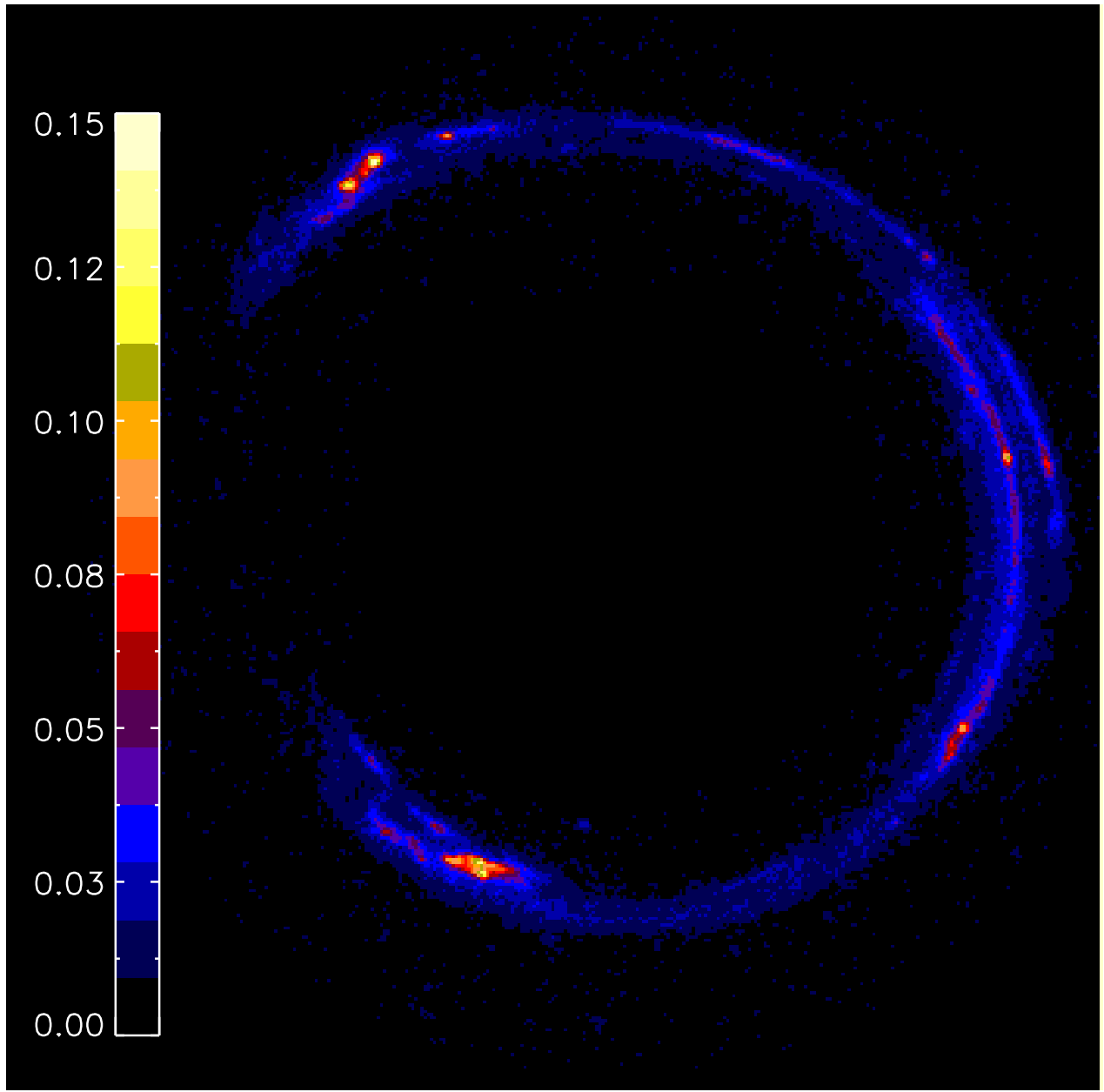}
  \caption{The HST WFC3 image of the cosmic horseshoe lens taken using the F475W filter.}
 \label{fig_1}
\end{figure}

\begin{figure}
 \includegraphics[width=\columnwidth]{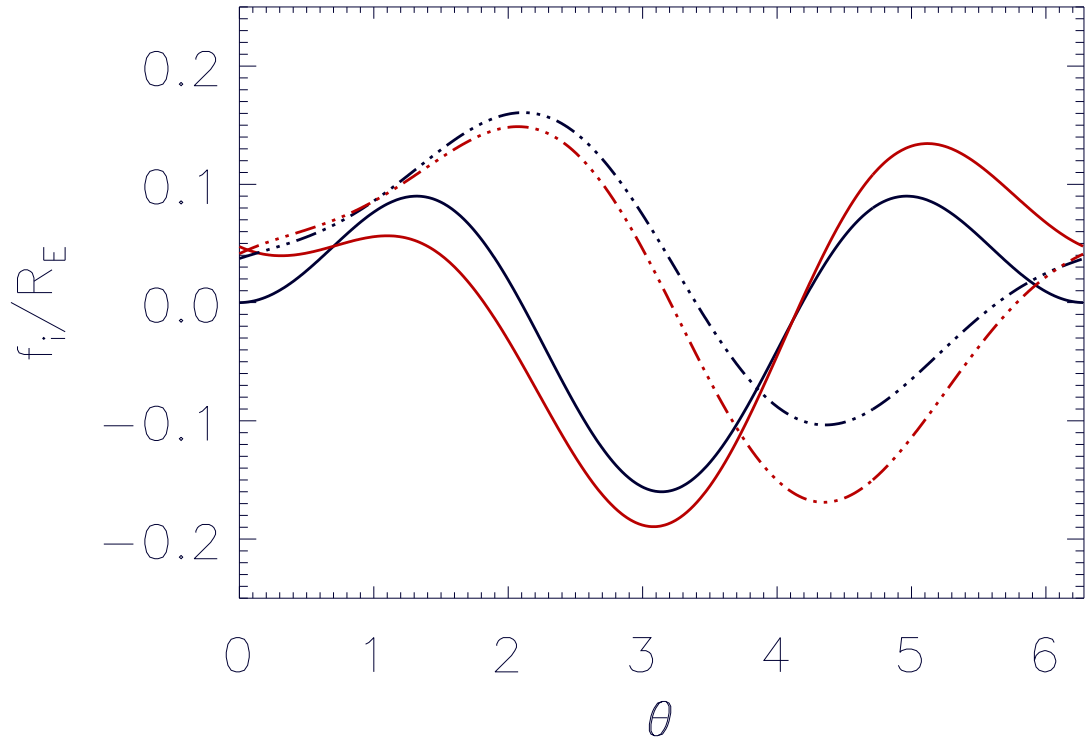}
  \caption{The initial guess (dark color) superimposed with the refined
 solution at Fourier order 2. The field $\frac{d f_0}{d \theta}$ is represented by a continuous line
while $f_1$ is represented with a dotted line.}
 \label{fig_2}
\end{figure}

The numerical fit of the data is performed on a selected area. First the area occupied by the arc is identified
by applying a threshold and then extending the area by performing a convolution with the pixels above the threshold. This
convolution procedure ensure that the outliers of the arc are properly included. Additionally a number of control
points (one third of the former points) are included by randomly selecting points in pure noise area in a large
ring around the arc. These control points are used to identify possible additional images that could be formed by the model
in dark areas. These additional points are also useful to control 
the statistics of errors in pure noise areas during the numerical minimization process. 
In the first step of the numerical refinement the fields are modeled by order 2 Fourier series. Given
a model of the fields a reconstruction of the source and images of the source are performed using the method
described is Sec. ~\ref{Source_rec} and ~\ref{Deconv}. 
Using the difference between the reconstructed images
and the actual images a weighted chi-square is estimated. The weights applied in the chi-square estimation are computed
using the noise expectation (see Sec. ~\ref{Sec_noise}). The reconstruction is applied first to the blue (F475W)
HST WFC3 calibrated image (see Fig. ~\ref{fig_1}). The main asset of the blue band is that the residual contribution of the central deflector
to the flux of the arc is very faint. Despite the weakness of the out-layers from the main
galaxy, a Sersic model was fitted to the deflector and the contribution around the arc was subtracted. The
typical amplitude of the subtraction is at the noise level. The refinement process is performed on
this processed image of the arc by using the Simplex method \cite{Nelder}. 
The approximate guess for the coefficients is used to initialize
the Simplex, and start the minimization process of the chi-square. It is important to note that the result of this
 minimization process does not produce significantly different results if the initial guess is changed.
Experiments were conducted by changing the parameters of the initial guess by an amplitude which is of the
order of the refinement. The results of these experiments indicates a general convergence to the same solution.
The stability of the solution is due to the perturbative approach. This method provides a fundamental reduction
of the degeneracy problem encountered in the modeling of gravitational lenses. The result of the Simplex minimization are
presented in Fig. ~\ref{fig_2}. In the next step the expansion in Fourier series of the numerical solution is extended
to higher order. First it is extended to order 3, the initial guess is the second order solution with zeros for the third
order terms. Similarly the solution is extended to order 4. An important point is the discussion of the noise and statistical
properties of the solutions at different orders.
\subsection{Noise and statistical properties of the solutions.}
\begin{figure}
 \includegraphics[width=\columnwidth]{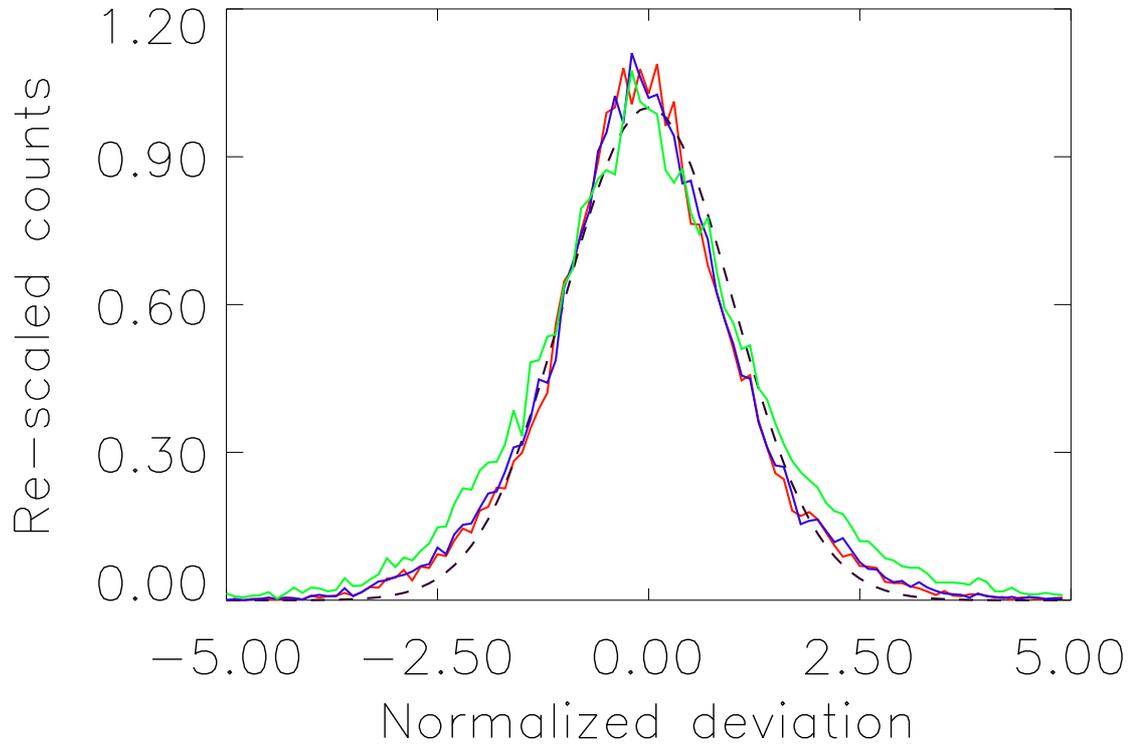}
  \caption{Histograms of the normalized deviations for the order 2 (green), order 3 (blue) and order 4 (red). The dashed 
line corresponds to the theoretical Gaussian expectation.}
 \label{hist1_fig}
\end{figure}

\begin{figure}
 \includegraphics[width=\columnwidth]{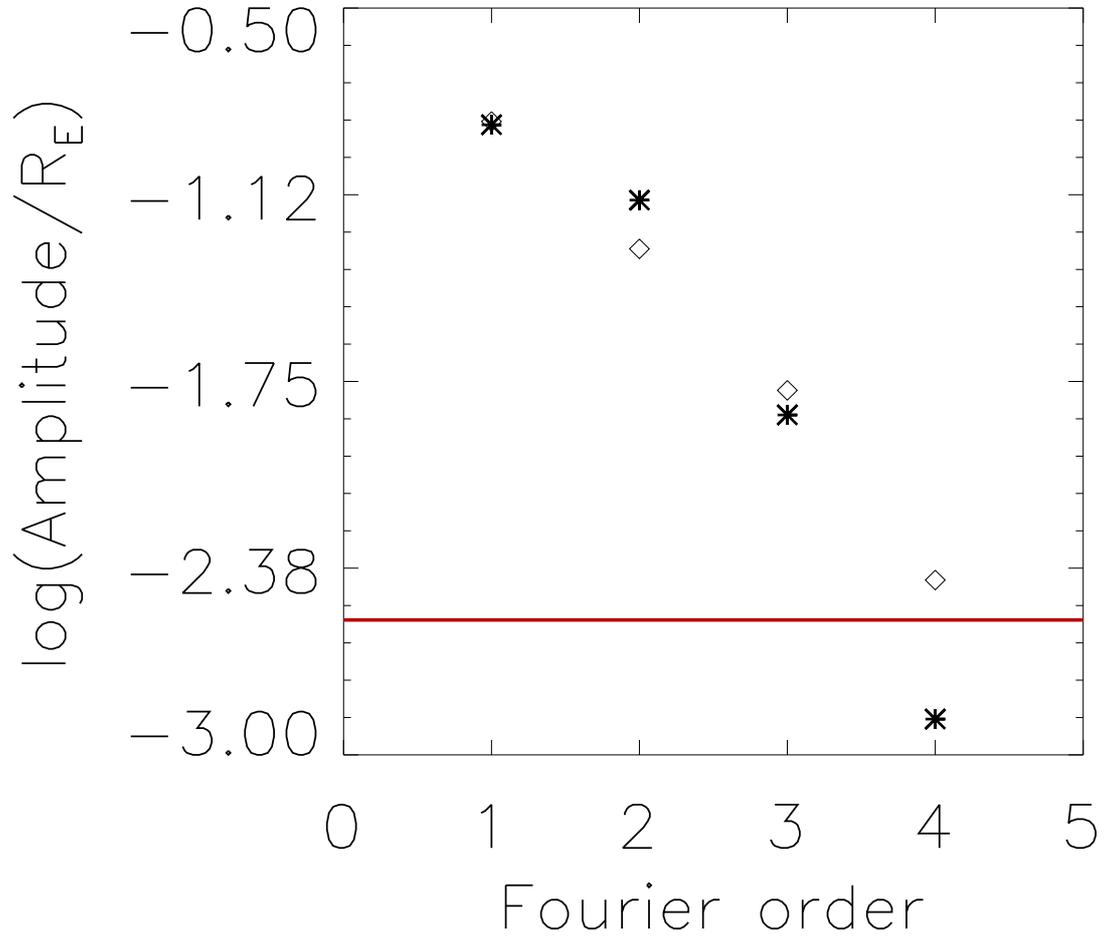}
  \caption{The amplitude of the Fourier components as a function of the order of the component. The red
line represents the $4 \sigma$ limit. The asterisks represents the components of $\frac{d f_0}{d \theta}$, while
the diamonds represents the components of $f_1$.}
 \label{fig_qx}
\end{figure}

The statistical expectation of the noise in the image $\sigma_0$ corresponds to the photon noise derived from
the photon counts. Using this statistical expectation for the noise the chi-square and $\chi^2$/dof are estimated for the different
models (see Table ~\ref{tab1}). The corresponding histogram of the normalized deviations for each model is presented
in Fig. ~\ref{hist1_fig}. For comparison the $\chi^2$/dof for the two other filters available in the HST archive 
are also presented in Table ~\ref{tab1}.
There is a very marked difference between the $\chi^2$/dof obtained at order 2 and order 3. The gain in $\chi^2$ obtained
by going to order 4 is much less spectacular. This suggests that the order 3 terms are important and significant. To test 
this hypothesis it is essential to estimate the amplitude of the noise fluctuation for the Fourier coefficients. The least-square
minimization is non-linear, but can always be linearized near the optimal solution. Let's define the optimal solution
$$
{\bf P}=[\alpha_{0,i},\beta_{0,i},\alpha_{1,i},\beta_{1,i}]_{\{i=1..4\}}
$$ 
We linearize the lens model $M({\bf P})$ near ${\bf P}$:
\begin{equation}
 M({\bf P+dP})= M({\bf P}) + \sum_n \frac{\partial M}{\partial p_n} dp_n
\label{linear_eq}
\end{equation}
Where $p_n$ and $dp_n$ are the components of the vectors $P$ and $dP$ respectively. The model in Eq. (~\ref{linear_eq})
is formally equivalent to a linear least-square with basis vectors, $\frac{\partial M}{\partial p_n}$. As a consequence
the errors on the parameters are directly the diagonal elements of the inverse of the corresponding normal least-square matrix.
Explicitly, the normal least-square matrix elements are,
$$
A_{ij}=\sum_{Image}  \frac{1}{\sigma_0^2} \frac{\partial M}{\partial p_i}  \frac{\partial M}{\partial p_j}
$$
Defining $C=A^{-1}$ the variance associated with parameter $p_n$ is, $\sigma_n=C_{nn}$. A numerical estimation of the matrix elements
$A_{ij}$ shows that the associated variance elements $\sigma_n$ are almost constant and equal to $0.5 10^{-3}$ in units of the Einstein
radius $R_E$. The calculation of the variance of the Fourier components allows a direct estimation of the significance of the components
at different orders (see Fig. ~\ref{fig_qx}). It is clear in Fig. ~\ref{fig_qx} that components of order $n \ge 3$ are well above
the $4 \sigma$ limit. This is a direct confirmation that the order 3 components are very significant and essential to the modeling of this lens. The amplitude of the order $4$ component is not significant for the $\frac{d f_0}{d \theta}$ field and is only marginally significant
for the $f_1$ field.
\label{Sec_noise}
\section{Analyzing the solution.}
\begin{figure}
 \includegraphics[width=\columnwidth]{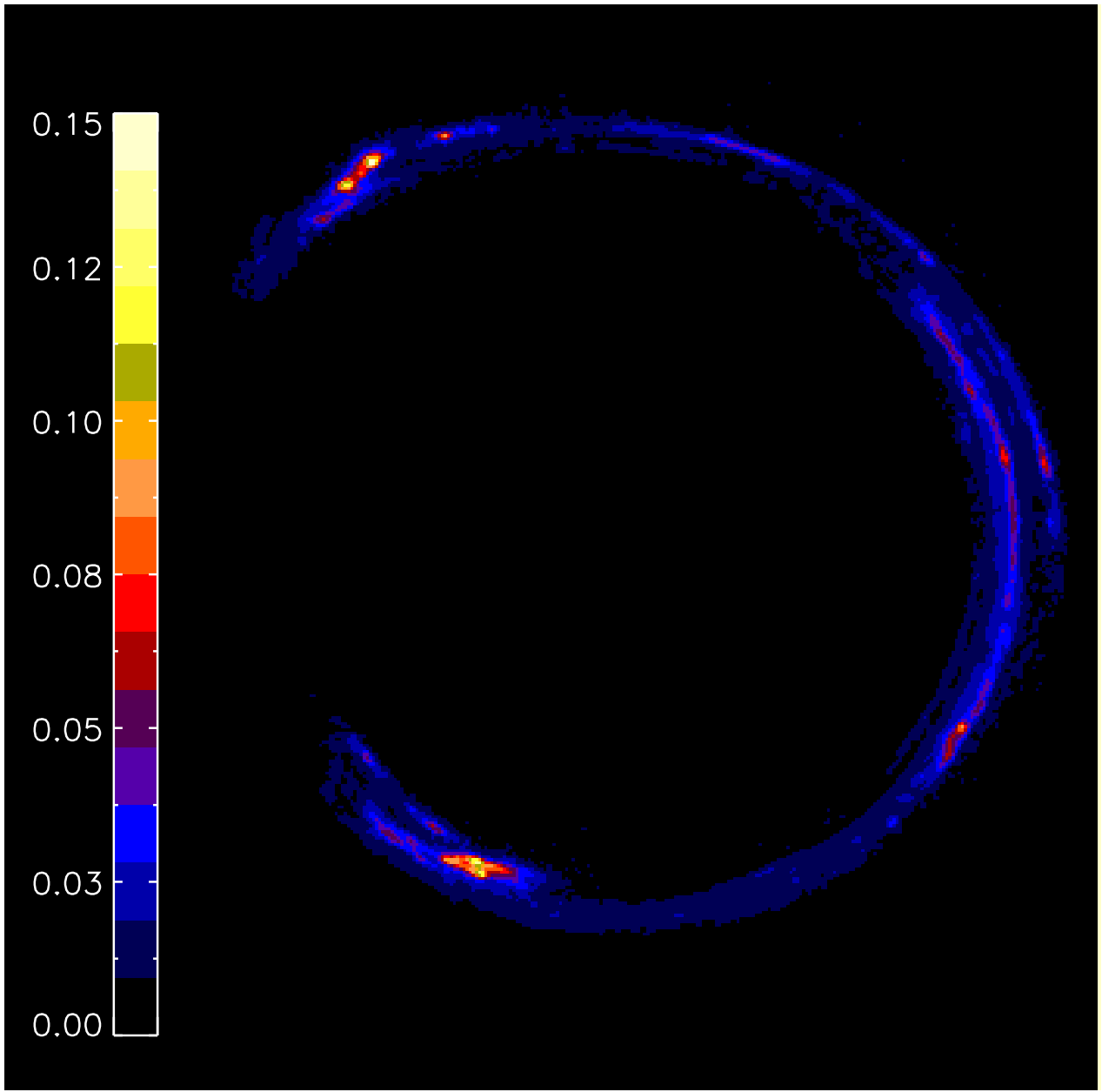}
  \caption{The reconstructed image in the F475W band.}
 \label{fig_im_r}
\end{figure}

\begin{figure}
 \includegraphics[width=\columnwidth]{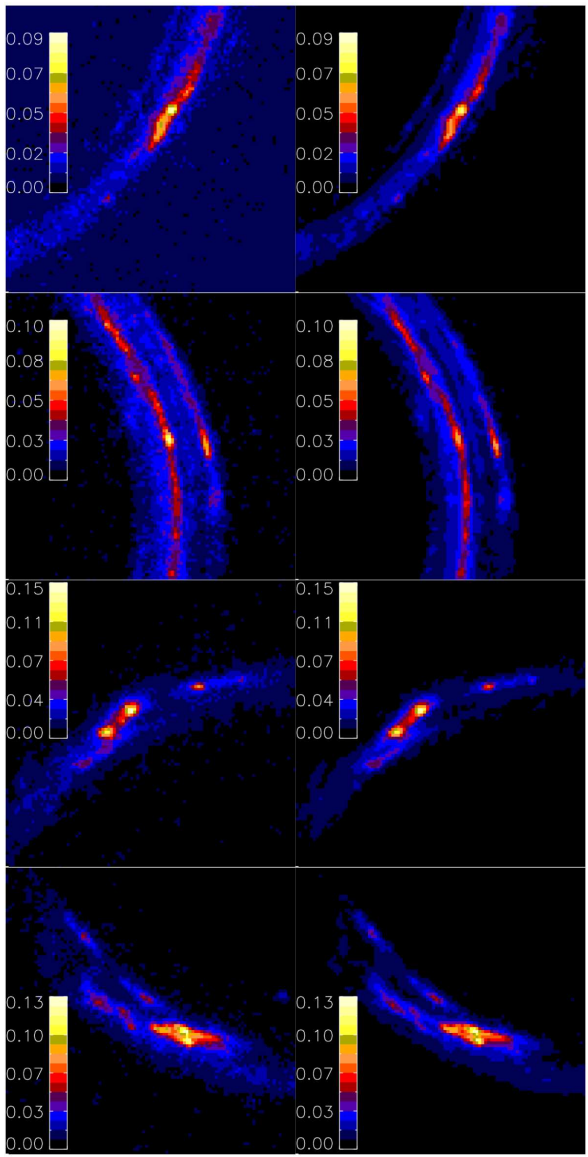}
  \caption{Details of the HST image in the F475W band (left) compared with the corresponding area
 in the reconstructed image (right).}
 \label{fig_detail}
\end{figure}

\begin{figure}
 \includegraphics[width=\columnwidth]{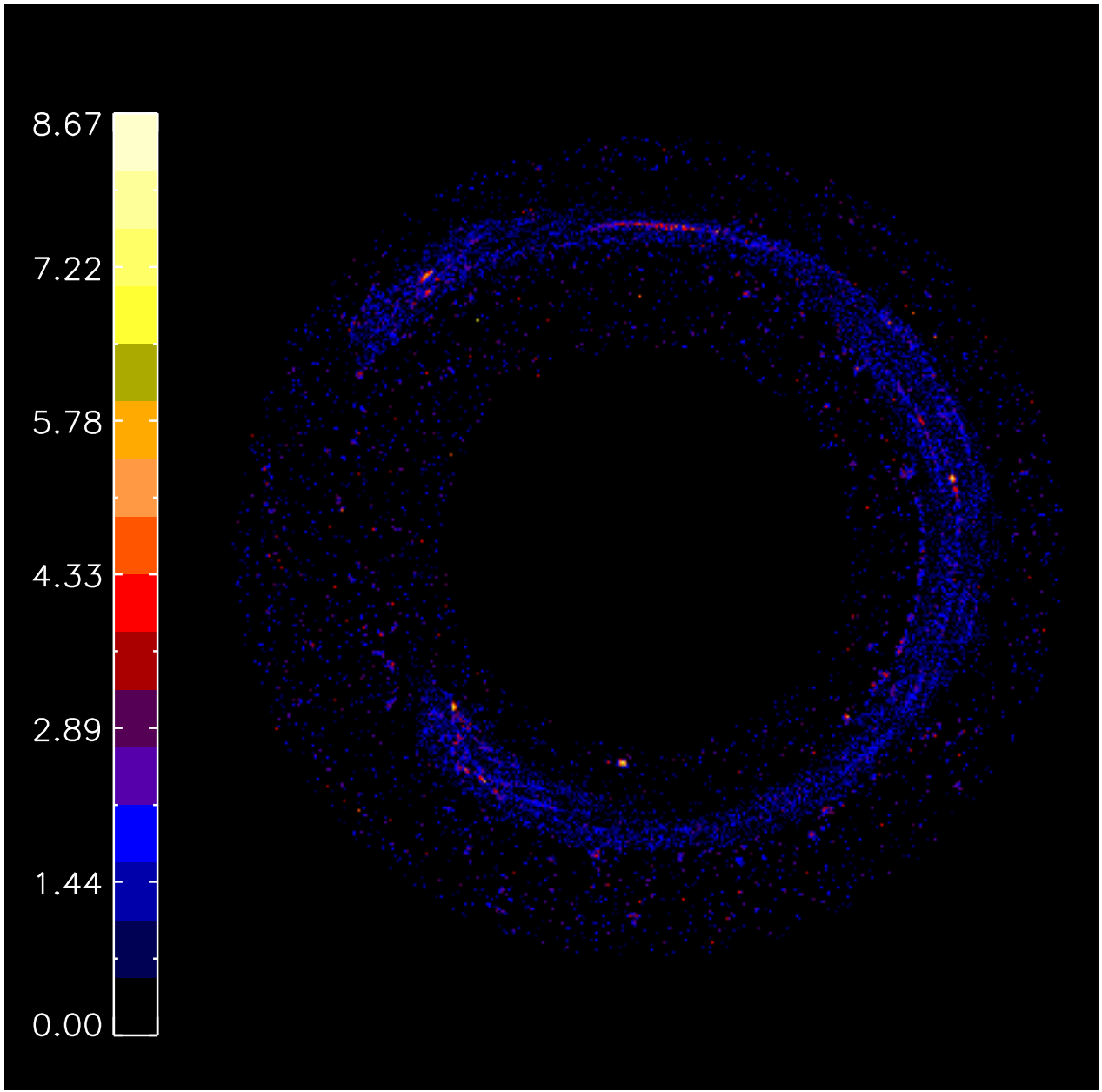}
  \caption{The absolute value of the difference between the HST image and the reconstructed image normalized by the
 noise expectation (F475w band). Note that additional points were added in pure noise area as control points (one third of total points).}
 \label{fig_diff2}
\end{figure}

\begin{figure}
 \includegraphics[width=\columnwidth]{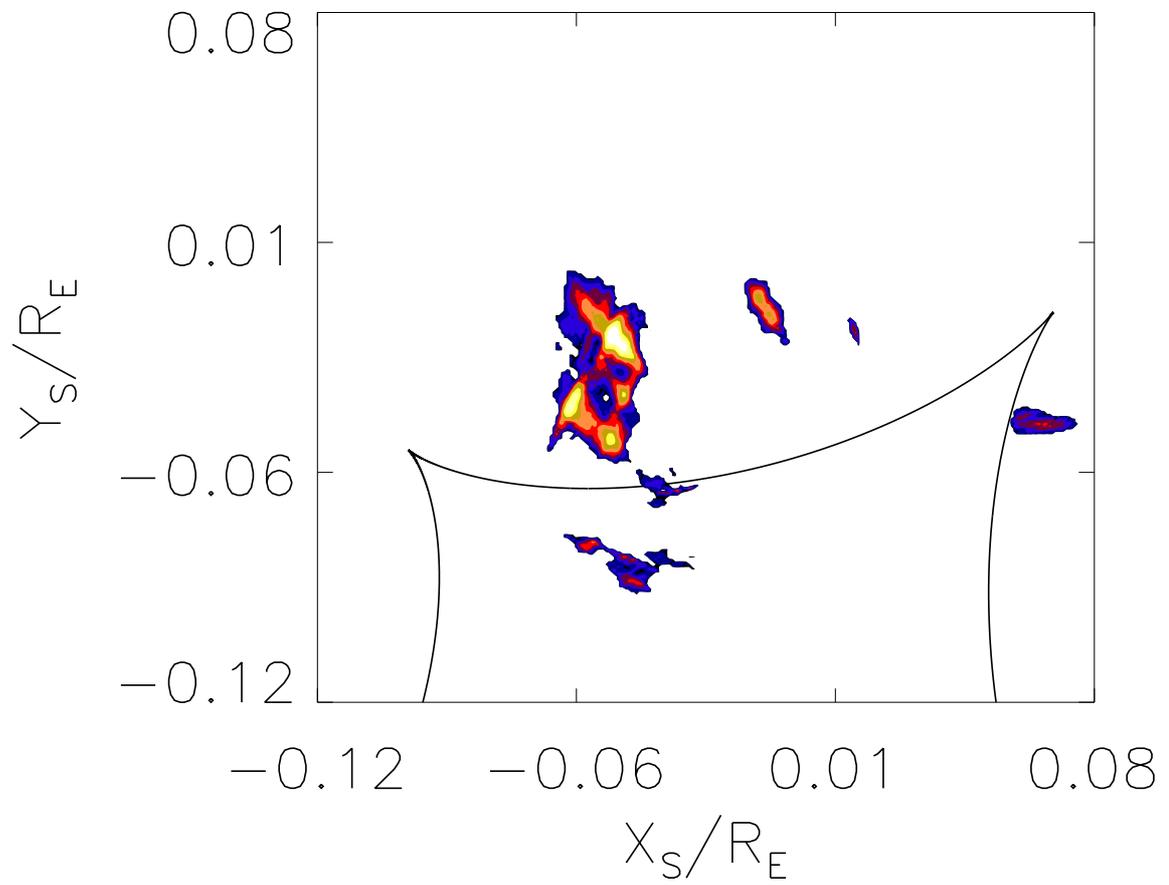}
  \caption{The reconstruction of the source superimposed with the caustic system of the lens.}
 \label{fig_caust}
\end{figure}

\begin{figure}
 \includegraphics[width=\columnwidth]{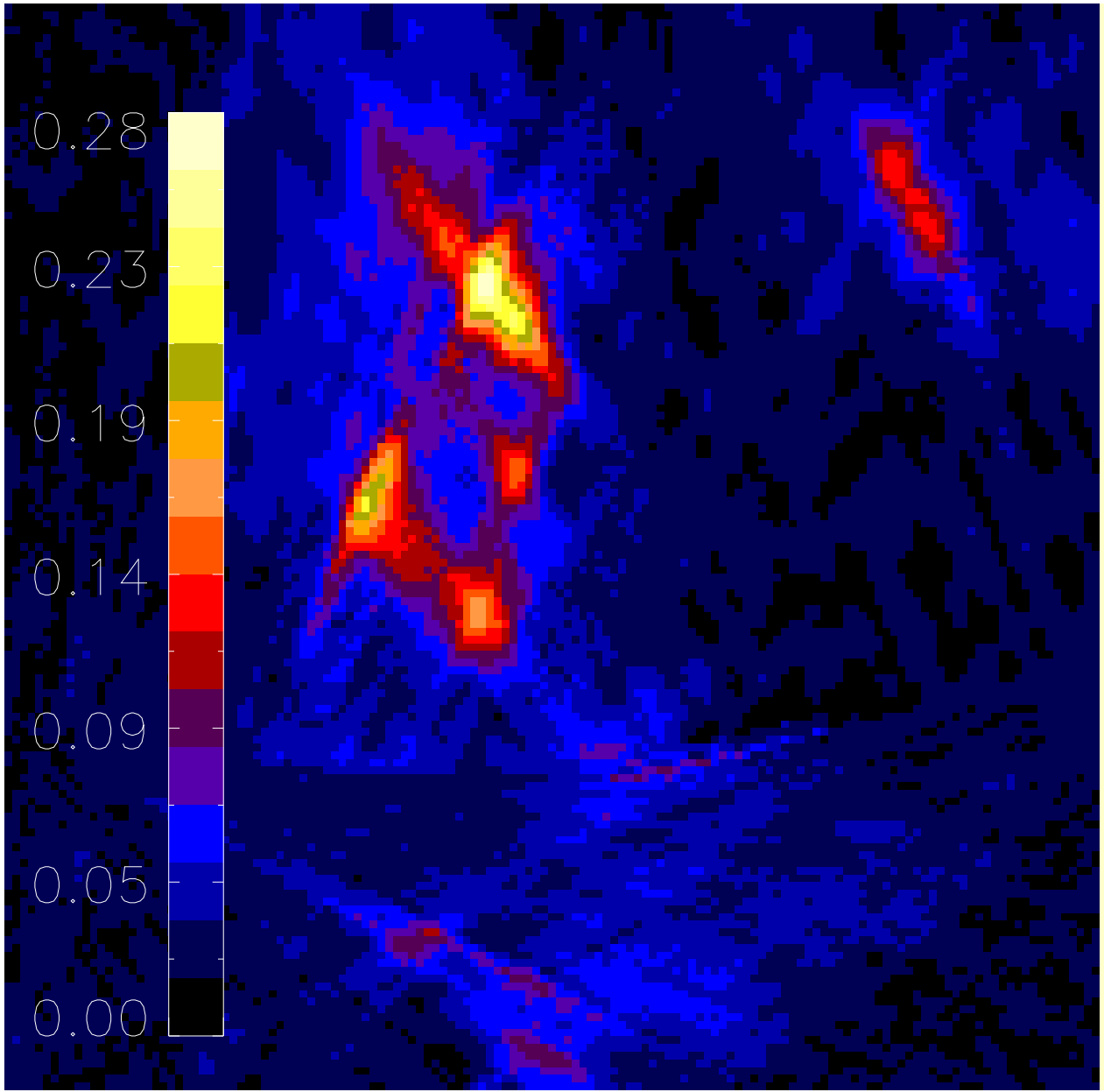}
  \caption{A detailed view of the source main component.}
 \label{fig_source}
\end{figure}

The structure of the fourth order solution is now explored in details. The reconstruction for the F475w filter
in the lens plane is presented in Fig. ~\ref{fig_im_r}. A comparison of the fine details of the solution and the original
HST image in the F475W band is presented in Fig. ~\ref{fig_detail}. A general appreciation of the quality of the
reconstruction is also provided by the difference image with the original image (see Fig. ~\ref{fig_diff2}). The reconstruction
in the source plane is presented in Fig. ~\ref{fig_caust}. For a more detailed view of the source see Fig. ~\ref{fig_source}.
\subsection{Effect of degeneracy induced by higher order perturbative terms.}
The first order perturbative expansion neglect the effect of higher order terms in the expansion.
For most gravitational arcs it is possible to reduce the higher order expansion to a first order expansion.
 Which means that some small degeneracy problem is present.
The amplitude of the correction due to higher order terms is evaluated using realistic (NFW)
models for the halo of the deflector. Let's consider a purely elliptical NFW halo which by definition has
no third order distortion of its isophotes. We consider the perturbative expansion of this elliptical NFW model
to perturbative order 2. The forced reduction of this expansion to order 1 introduces additional degenerate terms in the 
expansion. Could these terms be responsible for the order 3 terms that we observe in the lens model ?
The expansion at order 2 reads (~\cite{Alard2016}):
\begin{equation}
{\bf r_S} = \left( \kappa_2 \ dr -\kappa_3 \frac{dr^2}{2} -\tilde f_1-f_2 dr \right) {\bf u_r} -\left(\frac{d \tilde f_0}{d \theta}+\left(\frac{d f_1}{d \theta}-\frac{d f_0}{d \theta} \right) dr \right) {\bf u_{\theta}}
\label{eq_ordre2}
\end{equation}
$$
\kappa_3=\left[\frac{d^3 \psi_0}{dr^3}\right]_{r=1}
$$
And
$$
f_2=\frac{\partial^2 \psi}{\partial r^2}
$$
Assuming a thin arc model,
\begin{equation}
 dr=\frac{\tilde f_1}{\kappa_2}+\epsilon dr_2
 \label{eq_thin_exp}
\end{equation}
The arc presents in this lens present a thickness in the radial direction which is small with respect
to the general displacement fields. Thus Eq. ~\ref{eq_thin_exp} is certainly appropriate for an evaluation
of the amplitude of the third order terms. Introducing Eq. ~\ref{eq_thin_exp} in Eq. ~\ref{eq_ordre2} leads
to (~\cite{Alard2016}):
\begin{equation}
\tilde r_S=\left(\kappa_2 dr_2 - \kappa_3 \frac{\tilde f_1^2}{2 \kappa_2^2} -\frac{\tilde f_1 f_2}{\kappa_2} \right) {\bf u_r}-\left(\frac{d \tilde f_0}{d \theta}+\left(\frac{d f_1}{d \theta}-\frac{d f_0}{d \theta} \right) \frac{\tilde f_1}{\kappa_2} \right) {\bf u_{\theta}}
\label{eq_thin_exp2}
\end{equation}
Eq. ~\ref{eq_thin_exp2} is equivalent to Eq. ~\ref{pert_1} provided that the following subsitutions are performed:
\begin{equation}
\left\{
\begin{aligned}
 \tilde f_1 & \ \rightarrow  \tilde f_1+\kappa_3 \frac{\tilde f_1^2}{2 \kappa_2^2} +\frac{\tilde f_1 f_2}{\kappa_2} \\
 \frac{d \tilde f_0}{d \theta} &  \ \rightarrow   \frac{d \tilde f_0}{d \theta}+\left(\frac{d f_1}{d \theta}-\frac{d f_0}{d \theta} \right) \frac{\tilde f_1}{\kappa_2}
\end{aligned}
\right.
\label{pot_deg}
\end{equation}
To evaluate the amplitude of the additional terms in Eq. ~\ref{pot_deg} we use the NFW halo model.
The potential for a NFW halo reads (\cite{Meneghetti2003}):
\begin{equation}
\left\{
\begin{aligned}
 \phi(u)=\frac{1}{1-\ln(2)} g(u) \\
 u=\sqrt{\left( (1-\eta) x^2 + (1+\eta) y^2 \right)}
\end{aligned}
\right.
\label{pot_nfw}
\end{equation}
The parameter $\eta$ is related to the ellipticity of the halo. The potential normalization implies that the associated Einstein radius is equal to the typical halo size, which is a common situation for gravitational lenses. The definition of the function $g(u)$ reads:
\begin{equation}
g(u)=\frac{1}{2} \ln\left(\frac{u}{2}\right)^2+
\left\{
\begin{aligned}
& \ \ \ \  2 {\rm arctan}^2 \left (\sqrt {\frac{u-1}{u+1}} \right) &  \ \ u \ge 1\\
& -2 {\rm arctanh}^2 \left (\sqrt {\frac{1-u}{u+1}} \right) & \ \ u < 1
\end{aligned}
\right.
\label{g_def}
\end{equation}
Using the NFW potential (see ~\cite{Meneghetti2003} ) defined in Eq. ~\ref{pot_nfw} the functionals, $f_0$, $f_1$, and $f_2$ are calculated. The result
is introduced in Eq. ~\ref{pot_deg} to evaluate the correction due to the order 2 terms. For the $f_1$ field the correction is respectively
$\simeq 0.2 x_0 \eta$ and $\simeq 0.2 y_0 \eta$ for the $\cos(3 \theta)$ and $sin(3 \theta)$ terms. Where ($x_0,y_0$) are the source impact
parameters.The correction for $f_0$ is of smaller
amplitude. For this lens, $x_0 \simeq 0.03$, $y_0 \simeq0.13$, and $\eta \simeq 0.07$. As a consequence the largest third order
term is only of $\simeq 2 \ 10^{-3}$ which is similar to the $4 \sigma$ noise limit. Thus it is clear from this analysis
that the observed third order terms in this lens are not the consequence of neglected higher order terms in the perturbative expansion.
%
%
%
%
%
%
\section{Interpretation of the lens model.}
\begin{figure}
 \includegraphics[width=\columnwidth]{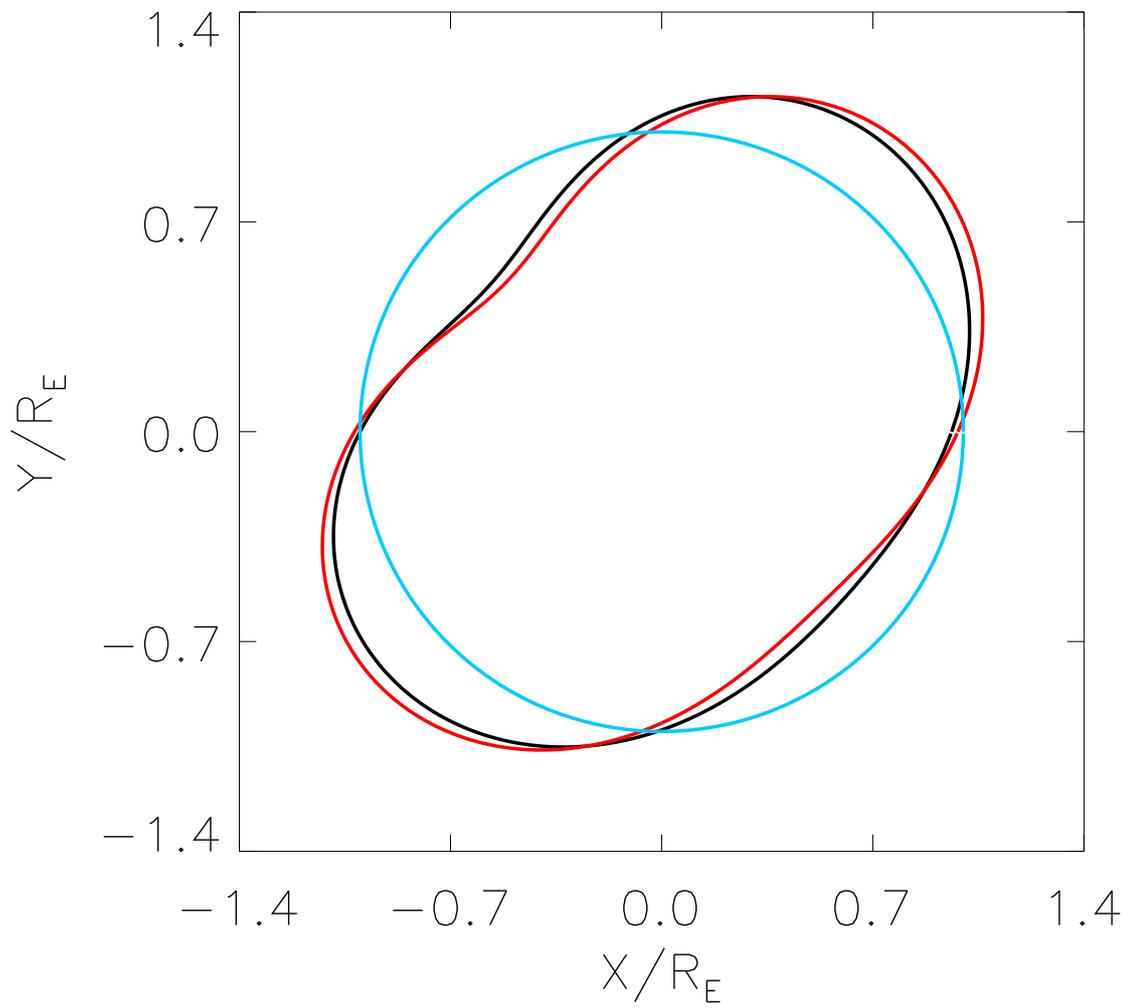}
  \caption{The potential iso-contours (red line) super-imposed with the iso-contours corresponding
to the outer distribution (black line). A circle with radius unity is plotted for reference (blue line).}
 \label{pot_contour}
\end{figure}

The reconstruction of the lens is directly related to the geometry of the potential. The potential iso-contours equation $dr=-f_0$
(~\cite{Alard2009}) is represented in Fig. ~\ref{pot_contour}. By relating the perturbative expansion to the multipole expansion (~\cite{Alard2009})
one can reconstruct the potential generated by the distribution of matter inside the Einstein circle (inner) and outside
the Einstein circle (outer). The iso-countour for the outer potential represented in Fig. ~\ref{pot_contour} is close to the potential
iso-contour. As a consequence most of the potential is generated outside the Einstein circle, and even more for the third order terms
where more than 90 \% of the potential originates in the outer distribution. The outer distribution includes a number of a galaxies
belonging to a small group of galaxies where the central deflector is the main element (~\cite{Belokurov2007}, ~\cite{Spiniello2011}, ~\cite{Agnello2013}). Thus it is interesting to evaluate
 the perturbating contribution of the accompanying galaxies in the group.  The first step is to identify the galaxies around the lens. A general
 search for galaxies in a radius corresponding to the size of the group ($\simeq 1$ arc minute ~\cite{Belokurov2007}) was performed. Objects
were identified by looking for local maxima's in a moving mesh with a size of 25 pixels. Punctual objects with corresponding width
not significantly larger than the PSF were eliminated. The Petrosian magnitude (~\cite{Petrosian}) is evaluated for the remaining objects in the different
photometric bands. The contribution of each galaxy to the lensing fields is estimated by assuming a proportionality relation between the red
(F875W) flux and the total mass. The potential of the perturbator is evaluated by considering the three following models (I) a spherical
isothermal sphere, (II) a point mass, (III) a spherical NFW profile. For each model the lensing fields $f_1$ and $\frac{d f_0}{d \theta}$ are 
evaluated using Eq. (~\ref{eq_f1_df0}). It is clear that the assumption of a proportionality between the mass and the red magnitude is crude.
However it is essential to note that about $90 \%$ of the contribution to the lensing perturbation is due to elliptical galaxies with
color similar to the central galaxy. Thus at least for this dominant sub-set all galaxies are quite similar
 and thus it is reasonable to assume a proportionality relation between flux and mass. The flux normalization factor is applied by dividing fluxes
by the flux of the main galaxy. As a result all masses are expressed in units of the main galaxy mass. In this reconstruction another source of error
come from the ellipticity of the potential. For an axis ratio of the ellipse equal to $1+\eta$ the typical percentage in error 
induced by the ellipticity on the fields $f_1$ and $\frac{d f_0}{d \theta}$ is of the order of $\eta$. For dark matter halos a typical value for $\eta$
is $\eta \simeq 0.2$. The uncertainty on mass is at least of about 10 \% since the flux to mass proportionality relation applies only to elliptical galaxies.
As a consequence an error of about 25 \% at least is expected for this kind of model. An estimation of the fields for the different models is presented
in Table ~\ref{tab2}. The point mass and NFW models prediction for the third order coefficients is in general agreement with the values reconstructed
from the lens model. As shown by the relative errors presented in Table ~\ref{tab3} the isothermal model exceeds the error expectation, while
the errors for two other models are consistent with the 20 to 30 \% relative error expectation. It is interesting to compare these results
with ~\cite{Dye2008} who could not find any contribution coming from the other group members. However ~\cite{Dye2008} used low resolution
images taken from the ground and could not reach the level of accuracy obtained with HST images. A more recent analysis of the cosmic horseshoe lens
was performed by ~\cite{Brewer2016} who could not find substantial evidence for substructures in the lens. However the problem of evaluating the contribution
of substructures in the lens is not equivalent to the evaluation of the group members contributions. The elements of the group are
more massive than substructures and on average are situated at larger distances than substructures. Thus it is not surprising that the ~\cite{Brewer2016}
analysis fails to identify the contribution of the group members. 
\section{Conclusion.}
The perturbative method allowed the reconstruction of the cosmic horseshoe lens without making any particular assumptions. The inclusion of third order terms
was dictated only by the necessity to optimize the chi-square. These third order terms were related to the group
contribution later in the analysis, but it was not necessary to make any hypothesis about the group contribution when reconstructing the lens.
In this case the group is made of quite a large number of galaxies and trying to make an extensive model including each object would require
too many parameters. Such models including many parameters are generally plagued with degeneracies issues which is a constant re-occuring problem
in conventional gravitational lens analysis. This analysis does not have to include all theses parameters but reduces the lens to a number
of fundamental parameters. It is clear that this minimal set of parameters (basically the expansion of the fields to order 3) corresponds
to the expectation of many models when the model include more parameters that the fundamental parameters (which would be the case here when modeling all the group). 
As a consequence it is clear that the perturbative approach is a method of choice for complex systems. The perturbative approach 
allows a model free, non-degenerate, fast and simple analysis of any gravitational lens system. It is important to note that even in the case of an a priori
simple lens system it is useful to apply the perturbative method since this method could reveal unexpected complex contributions. Essentially in the same way
that the contribution of the group was discovered without making any initial hypothesis about the presence of the group.
\section*{Acknowledgements}
This work is based on observations made with the NASA/ESA Hubble Space Telescope, obtained from the data archive at the Space Telescope Science Institute. STScI is operated by the Association of Universities for Research in Astronomy, Inc. under NASA contract NAS 5-26555.
\begin{table}
	\centering
	\caption{The $\chi^2$/dof as a function of the order of the Fourier series
expansion of the fields for the three different filters.}
\begin{tabular}{llll} 
  Order & 2 & 3 & 4 \\
  \hline
    $\chi^2$/dof (F475W) & 2.29 & 1.38 & 1.28 \\
\hline
    $\chi^2$/dof (F675W) & 1.88 & 1.32 & 1.22 \\
\hline
    $\chi^2$/dof (F875W) & 2.14 & 1.52 & 1.47 \\
\end{tabular}
\label{tab1}
\end{table}
\begin{table}
	\centering
	\caption{Models of the perturbation due to the group members.}
\begin{tabular}{lllll} 
  Model & $\frac{d f_0}{d \theta};\cos(3 \theta)$ & $\frac{d f_0}{d \theta};\sin(3 \theta)$ & $f_1;\cos(3 \theta)$ & $f_1;\sin(3 \theta)$ \\
  \hline
   Isothermal & 0.002 &   -0.0073 & 0.0071 &  0.00175 \\
   Point mass & 0.01 &   -0.0123  &  0.0126 &  0.01 \\
   NFW        & 0.0094  &  -0.011 &  0.011  & 0.0096 \\
  \hline
  Reconstruction & 0.011  & -0.0078 &  0.0089  & 0.014 
\end{tabular}
\label{tab2}
\end{table}
\begin{table}
	\centering
	\caption{Relative deviation from reconstructed coefficients for different model of perturbator
potential.}
\begin{tabular}{lll} 
  Model & $\frac{d f_0}{d \theta}$ & $f_1$ \\
  \hline
   Isothermal & 0.68 & 0.75 \\
   Point mass & 0.36 & 0.32 \\
   NFW        & 0.26 & 0.3 \\
\end{tabular}
\label{tab3}
\end{table}
\begin{figure}
 \includegraphics[width=\columnwidth]{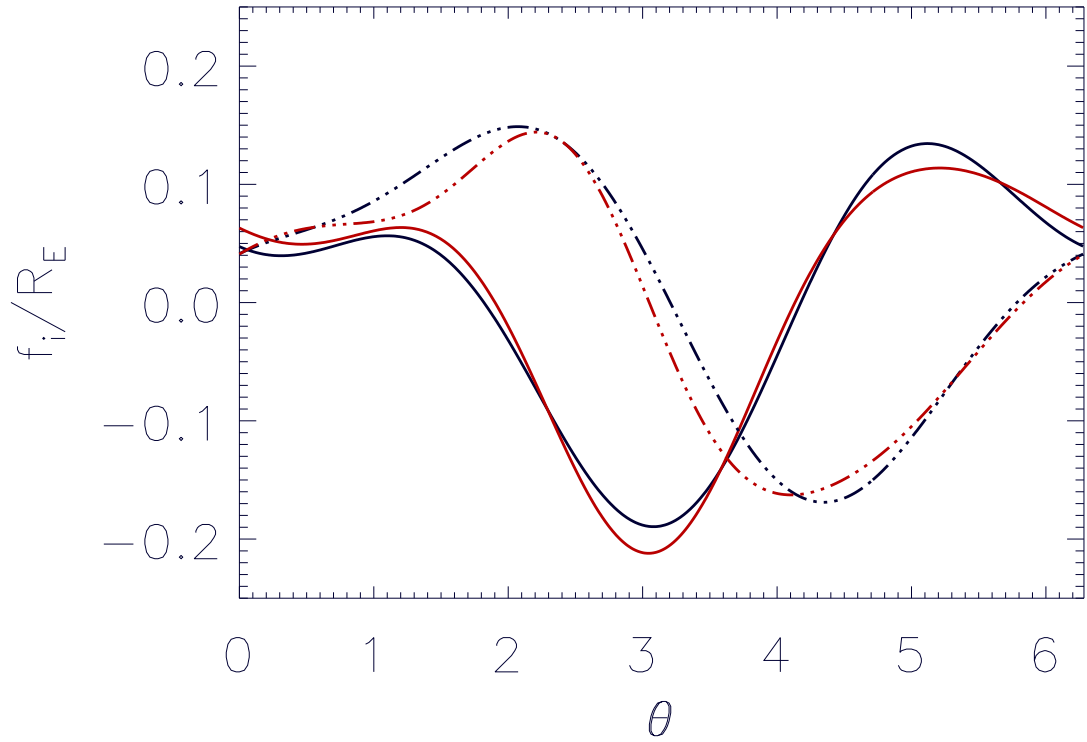}
  \caption{The solution for the fields at order 2 (dark color) superimposed with the refined
 solution at Fourier order 4. The field $\frac{d f_0}{d \theta}$ is represented by a continuous line
while $f_1$ is represented with a dotted line.}
 \label{fig_3}
\end{figure}
\end{document}